\documentclass[12pt]{article}
\usepackage[margin=0.5in]{geometry}
\usepackage{amsmath,amssymb}
\usepackage{hyperref}
\hypersetup{
  pdftitle={Astrocytic resource diffusion stabilizes persistent activity in neural fields},
  colorlinks=true,
  linkcolor=black,
  citecolor=black,
  urlcolor=blue
}
\usepackage{cite}
\usepackage{graphicx}
\usepackage{microtype}
\usepackage{authblk}
\usepackage{lmodern}
\usepackage[T1]{fontenc}
\usepackage{titlesec}
\usepackage{lmodern}
\usepackage[T1]{fontenc}
\usepackage{titlesec}
\usepackage{authblk}
\usepackage{abstract}

\titleformat{\section}{\large\bfseries\sffamily}{\thesection}{1em}{}
\titleformat{\subsection}{\normalsize\bfseries\sffamily}{\thesubsection}{1em}{}
\titleformat{\subsubsection}{\normalsize\itshape\sffamily}{\thesubsubsection}{1em}{}

\title{\sffamily\LARGE\bfseries Astrocytic resource diffusion stabilizes persistent activity in neural fields}

\usepackage{abstract}

\title{\textbf{Astrocytic resource diffusion stabilizes persistent activity in neural fields}}

\author[1]{Noah Palmer}
\author[2]{Heather L. Cihak}
\author[3]{Daniele Avitabile}
\author[1]{Zachary P. Kilpatrick}

\affil[1]{Department of Applied Mathematics, University of Colorado Boulder, Boulder, CO}
\affil[2]{School of Mathematics, University of Minnesota, Minneapolis, MN}
\affil[3]{Department of Mathematics, Vrije Universiteit Amsterdam, Amsterdam, The Netherlands}

\date{}

\begin{document}
\maketitle

\begin{abstract}
Persistent neural activity underlying working memory requires sustained synaptic transmission, yet the metabolic and neurotransmitter support provided by astrocyte networks is largely absent from spatially extended neural circuit models. We introduce a coupled astrocyte–neural field model in which synaptic efficacy is regulated by depletion and recovery of a conserved resource pool recycled and spatially redistributed through diffusively coupled astrocytes. We obtain explicit stationary bump profiles and self-consistency conditions for bump width and amplitude on a canonical ring architecture. Linearizing about these solutions while carefully accounting for perturbations at bump boundaries, we analyze the resulting spectral problem governing stability. Our analysis, supported by numerical simulations and low-dimensional Fourier truncations, reveals a two-stage stabilization mechanism: astrocytic diffusion smooths resource asymmetries created by small bump displacements, and synaptic replenishment transfers this smoothing back to the synaptic pool. Together, sufficiently strong diffusion and replenishment suppress drift instabilities and enlarge the parameter regime in which stationary bumps persist.

\medskip\noindent\textbf{Relevance to Life Sciences.} Astrocytes recycle neurotransmitters and redistribute ions and metabolic substrates through diffusively coupled networks, yet existing spatially extended neural circuit models rarely account for this resource cycling. As a result, there are few frameworks for studying how the interplay between local synaptic depletion and nonlocal astrocytic transport shapes spatiotemporal neural dynamics. We address this gap with a model whose conserved resource structure ensures that sustained firing in one region draws resources from neighboring tissue through the astrocyte network, mirroring observed metabolic support patterns. Our results suggest that both diffusive coupling strength and neurotransmitter recycling rates are critical for maintaining stable persistent activity, offering testable predictions linking glial network features to working memory robustness.

\medskip\noindent\textbf{Mathematical Content.} We derive explicit stationary bump solutions to the coupled astrocyte–neural field system on a periodic domain with cosine connectivity and Heaviside firing rate. Self-consistency conditions yield bump half-width and amplitude through a nonlinear relation for the resource depletion level. Linearization produces a piecewise-smooth spectral problem whose coefficients, determined by the interaction of the Heaviside nonlinearity with the resource coupling, depend on the sign of the perturbation at each bump boundary, generating four distinct cases. For each case, we construct an Evans function whose real zeros determine linear stability. The astrocyte diffusion equation is solved explicitly, yielding constant-coefficient ordinary differential equations on three subdomains. We analyze the limiting regimes of zero and infinite diffusion, obtaining closed-form expressions for the linearized resource perturbations. Analytic predictions are validated against numerical simulations and a low-dimensional Fourier truncation of the perturbed system.
\end{abstract}

\section{Introduction}
\label{sec:intro}

Working memory, the ability to hold and manipulate information over short time scales, has long been linked to persistent neural activity in which neurons sustain firing during and after a transient sensory input~\cite{Ma,Funahashi}. In oculomotor delayed response tasks, electrophysiological recordings from macaque prefrontal cortex reveal sustained, location-specific firing throughout the delay period following stimulus removal~\cite{Goldman,Constantinidis,Funahashi,Wimmer}. At the population level, remembered stimulus features are encoded in low-dimensional subspaces that remain stable across the delay even as single-neuron activity varies strongly~\cite{Murray2017,Spaak2017}, and trial-to-trial stochastic drift of population activity predicts memory errors~\cite{Compte,Wimmer}.

These observations motivate models in which working memory representations take the form of spatially localized regions of elevated firing, commonly referred to as bump attractors~\cite{Wang}, whose peak location encodes a remembered continuous stimulus feature such as spatial location, orientation, or color. In primates, support for bump-like dynamics comes from spatially tuned persistent firing and trial-to-trial correspondence between bump drift and saccade errors in prefrontal cortex~\cite{Goldman,Constantinidis,Funahashi,Wimmer}, attractor dynamics predicting categorical working memory judgments~\cite{mahrach2024cholinergic,thrower2023decoding}, and interactions between persistent activity and activity-silent synaptic traces underlying serial dependence~\cite{barbosa2020interplay}. Ring attractor dynamics have also been directly observed in the Drosophila head direction system~\cite{kim2017ring}, and human neuroimaging links sustained fronto-parietal activity to working memory capacity~\cite{Constantinidis}. Beyond their biological relevance, bump attractor models are mathematically tractable while capturing neurobehavioral dynamics of interest, making them a canonical modeling framework.

Neural field models describe the spatiotemporal evolution of population activity using integro-differential equations and are a common framework for analyzing bump attractor dynamics~\cite{Amari,Bressloff2012}. A key advantage of these models is that their parameters often retain biophysical meaning, so that changes in bump stability or dynamics can be traced directly to biological substrates. Laterally inhibitory connectivity gives rise to stationary bump solutions representing persistent, localized activity~\cite{Pinto,Coombes2003}. Incorporating negative feedback into these models reveals a rich landscape of instabilities: the destabilization of stationary bumps, traveling waves, and oscillatory breathers~\cite{Pinto,Folias04,coombes2005bumps,shusterman2008baseline}. In kind, physiologically-inspired models incorporating short term depression shapes bump motion and robustness over long time scales~\cite{Kilpatrick10,Cihak2024}, but typically treat synaptic resources as locally consumed without replenishment from an external pool. Synaptic resource cycling via astrocytes has been considered in space-free settings~\cite{Pitta,Kozachkov2025}, but has not been incorporated into spatially extended neural circuit models capable of capturing the interplay between local depletion and nonlocal transport.

In biological circuits, released neurotransmitter diffuses across the synaptic cleft and binds to postsynaptic receptors, but eventually undocks and is taken up by surrounding astrocytes for recycling (Fig.~\ref{fig:schematic}A,~\cite{Schousboe13,Weber2015}). Recycled resources can then be returned to the originating synapse or trafficked through gap junction-coupled astrocyte networks to other synapses across brain tissue~\cite{Giaume2010,Pannasch2013}. 
Astrocytes thus have a well-established role in synaptic plasticity, learning, and memory \cite{Ota2013,Williamson2025,Kozachkov2025}, and while astrocyte-mediated transmission supports persistent activity in neuron–glia models~\cite{Pitta}, existing work has largely ignored the spatiotemporal dynamics of such interactions.

To address this gap, we introduce a coupled astrocyte-neural field model in which synaptic efficacy is regulated by depletion and recovery of a conserved resource pool that is replenished and spatially redistributed through diffusively coupled astrocytes. We derive explicit stationary bump solutions, analyze their linear stability through an Evans function construction, and show that astrocytic diffusion stabilizes bumps against drift instabilities, enlarging the parameter regime in which persistent localized activity is maintained. Readers seeking background and further reading on bump attractor models, astrocytic resource recycling, and the Evans function stability framework are referred to Appendices~\ref{app:bg} and~\ref{app:read}.

\section{Astrocyte-Neural Field Model}
\label{sec:system}

\begin{figure}[t!]
    \centering
    \includegraphics[width=0.95\linewidth]{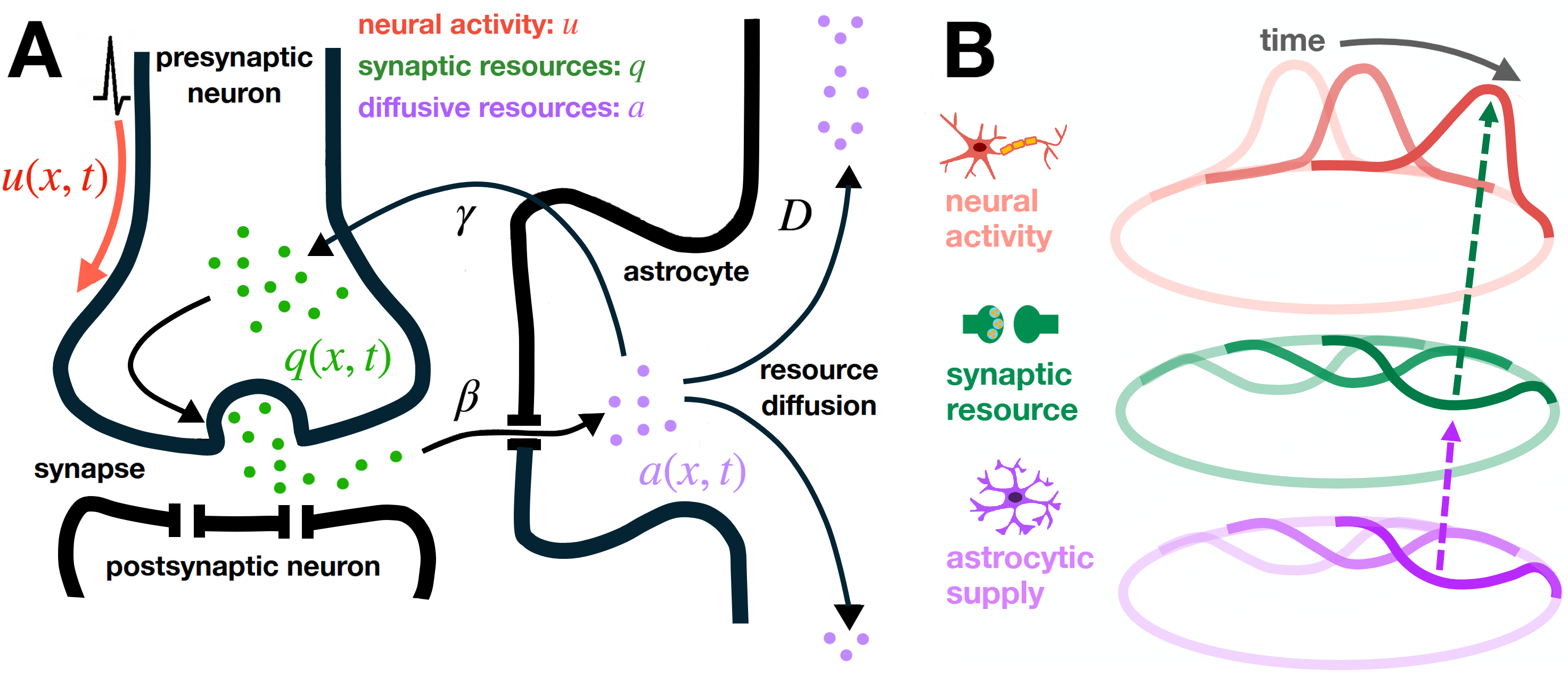}
    \caption{{\bf Conceptual schematic of the coupled astrocyte--neural field model.}
{\bf A.}~Local neuron--astrocyte interaction. Neural activity $u(x,t)$ depletes presynaptic resources $q(x,t)$ at rate $\beta$, which are replenished from astrocytic resources $a(x,t)$ at rate $\gamma$. Astrocytic resources are spatially redistributed through a diffusively coupled astrocyte network with diffusion coefficient $D$.
{\bf B.}~Neural field interpretation. Spatially localized neural activity depletes synaptic resources (green dashed arrow), while astrocytic diffusion redistributes resources across space (purple dashed arrow), stabilizing persistent bump attractors over time. }
    \label{fig:schematic}
\end{figure}

We consider a one-dimensional neural field model defined on the ring $x\in[-\pi,\pi]$ with periodic boundary conditions. The model describes the activity of a spatially distributed neuronal population, with synaptic efficacy regulated by depletion and recovery of a conserved resource recycled through a diffusively coupled astrocytic network (Fig.~\ref{fig:schematic}). The governing equations are
\begin{align}
\frac{\partial u}{\partial t}(x,t) &= -u(x,t)+\int_{-\pi}^\pi w(|x-y|)q(y,t)f(u(y,t))\ dy,\nonumber\\
\frac{\partial q}{\partial t}(x,t) &= \gamma a(x,t)\bigl(1-q(x,t)\bigr)-\beta f(u(x,t))q(x,t),\label{eq:anfmodel}\\
\frac{\partial a}{\partial t}(x,t) &= \beta f(u(x,t))q(x,t)-\gamma a(x,t)\bigl(1-q(x,t)\bigr)
+ D\frac{\partial^2 a}{\partial x^2}(x,t).\nonumber
\end{align}
Here $u(x,t)$ represents the mean activity of neurons at position $x$ and time $t$. The function $f(u)$ is the firing-rate nonlinearity which maps neural activity to population firing output. The synaptic connectivity kernel $w(|x-y|)$ describes the spatial structure of synaptic interactions. We will assume $w(|x-y|)$ is laterally inhibitory, meaning neurons with a similar stimulus preference excite each other and neurons with dissimilar preferences are inhibitory, a feature observed in many sensory systems. The variables $q(x,t)$ and $a(x,t)$ represent presynaptic and astrocytic resource availability, respectively.

Rather than modeling a specific biochemical substrate, the resource variables are used as proxies for multiple metabolic and neurotransmitter pathways~\cite{tsodyks1998neural}. 
Presynaptic resources are depleted at a rate $\beta f(u)$ by neuronal firing and transferred to the astrocytic resource pool. Astrocytic resources $a(x,t)$ replenish presynaptic resources $q(x,t)$ at rate $\gamma$, depending on the availability of both pools, and diffuse spatially through the astrocyte network with diffusion coefficient $D$.
A timescale could in principle be introduced for the resource variables separately from the membrane timescale governing $u$; here we absorb it into the rates $\beta$, $\gamma$, and $D$ by working in units of the membrane time constant. The depletion rate $\beta$ corresponds to the inverse recovery timescale of short-term synaptic depression, which is experimentally observed to lie in the range $100$--$1000\,\text{ms}$~\cite{tsodyks1997neural,abbott1997synaptic}, giving $\beta\sim 0.01$--$0.1$ in our units. The replenishment timescale is set by $(\gamma \cdot A_0)^{-1}$, where $A_0$ is the astrocytic resource level at steady state. For the parameter regimes we consider this falls in a similar range, consistent with observed glutamate recycling rates through the glutamate--glutamine cycle~\cite{Schousboe13,depitta2016astrocytes}.
The total amount of resource, $R_{\rm tot}$, is conserved, meaning that
\begin{align*}
    \frac{1}{2\pi}\int_{-\pi}^\pi\left(q(y,t)+a(y,t)\right)\,dy = R_{\rm tot}.
\end{align*}
The conservation constraint fixes the mean resource level but does not otherwise affect the existence or stability of stationary solutions.
This allows us to normalize $R_{\rm tot}=1$ and proceed with an explicit analysis of stationary bump solutions.

\section{Construction of stationary bump solutions}
\label{sec:bumpexistence}

To facilitate explicit analysis, we focus on a canonical setting.
Setting $R_{\rm tot}=1$, $f(u)=H(u-\theta)$, and selecting a laterally inhibitive kernel $w(x)=\cos(x)$ allows us to analytically characterize stationary bump solutions and highlights the role of astrocyte-mediated resource recycling. A stationary solution $u(x,t)= U(x)$, $q(x,t)=Q(x)$ and $a(x,t)=A(x)$ satisfies
\begin{align*}
U(x) &= \int_{-\pi}^\pi \cos(x-y)\,Q(y)\,H(U(y)-\theta)\,dy,\\
0 &= \gamma A(x)\bigl(1-Q(x)\bigr)
     - \beta\bigl[H(x+\Delta)-H(x-\Delta)\bigr]Q(x),\\
0 &= \gamma A(x)\bigl(1-Q(x)\bigr)
     - \beta\bigl[H(x+\Delta)-H(x-\Delta)\bigr]Q(x)
     + D A''(x),
\end{align*}
where we have used the fact that $H(U(x)-\theta)=H(x+\Delta)-H(x-\Delta)$ for a symmetric bump with $\Delta$ the bump's half-width.
By translation invariance, we may assume that the bump is centered at the origin, with $U(\pm\Delta)=\theta$, $U(x)>\theta$ for $x\in(-\Delta,\Delta)$, and $U(x)<\theta$ otherwise.
The corresponding synaptic and astrocytic resource profiles satisfy
\begin{align*}
    A(x) = A_0~~(\text{constant}),\quad Q(x) = \frac{\gamma A_0}{\beta  H(U(x)-\theta)+\gamma A_0}.
\end{align*}
Thus, $Q$ takes on distinct constant values within the active region and outside of it
\begin{align*}
    Q(x)=\begin{cases}
        c_0,& |x|<\Delta\\
        1,&  |x|\ge \Delta
    \end{cases}, \quad c_0:= \frac{\gamma A_0}{\beta+\gamma A_0}\in (0,1).
\end{align*}
The neural activity variable can be solved for explicitly as
\begin{align*}
    U(x) = c_0\int_{-\Delta}^\Delta \cos(x-y) \ dy = 2c_0\sin(\Delta)\cos(x).
\end{align*}
The constant $A_0$ is determined by the global resource balance
\begin{align}
    2\Delta c_0+2(\pi-\Delta)= 2\pi(1-A_0)\implies A_0=\frac{\Delta}{\pi}(1-c_0). \label{eq:A0}
\end{align}
Eliminating $A_0$ yields a quadratic equation for $c_0$
\[
\frac{\gamma\Delta}{\pi}c_0^2-\left(\beta+\frac{2\gamma\Delta}{\pi}\right)c_0+\frac{\gamma\Delta}{\pi}=0,
\]
which admits a unique solution $c_0\in(0,1)$ for each $\Delta$:
\begin{align}
c_0(\Delta)=\frac{\pi\beta +2\pi\gamma \Delta-\sqrt{\pi^2\beta^2+4\pi\beta\gamma \Delta}}{2\gamma \Delta}.\label{eq:c0}
\end{align}
The self-consistency condition $U(\pm\Delta)=\theta$ gives an implicit expression for the half width of the bump in terms of the other parameters
\begin{align}
    \theta = c_0(\Delta)\sin(2\Delta).\label{eq:thetadelta}
\end{align}

\begin{figure}[t!]
    \centering
    \includegraphics[width=0.8\linewidth]{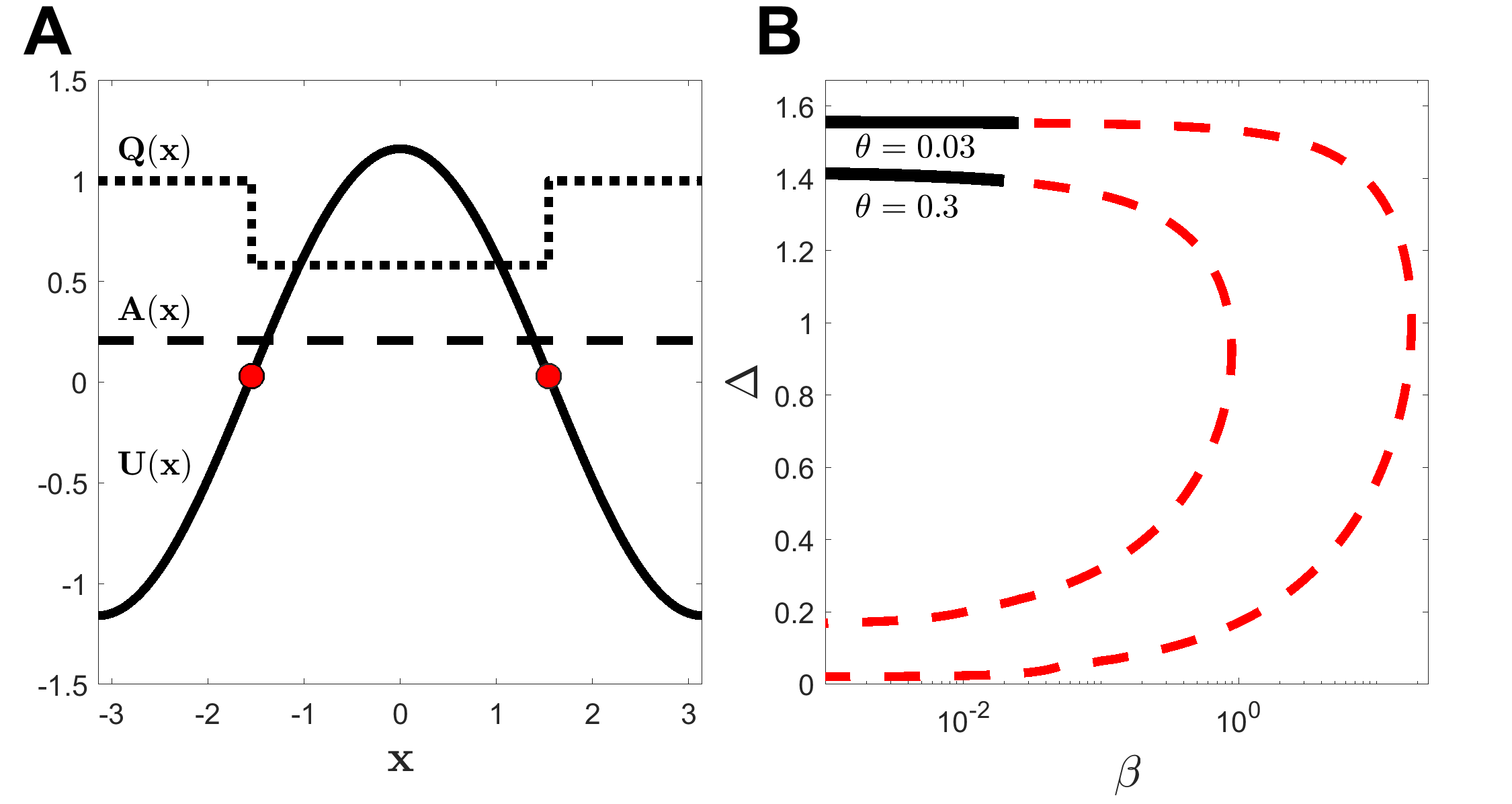}
    \caption{{\bf Stationary bump solutions and width dependence on synaptic depletion.} {\bf A.} Example stationary bump solution $(U,Q,A)$ in the cosine-kernel, Heaviside-rate case, showing the active region boundaries at $x=\pm\Delta$.
    Parameters: $(\beta,\theta)=(0.1,0.03)$.
    {\bf B.} Bump half-width $\Delta$ as a function of synaptic depletion strength $\beta$ for several threshold values $\theta$.
    Solid black (resp., dashed red) curves denote parameter regimes in which the linear stability analysis predicts the corresponding stationary bump to be stable (resp., unstable).
    Other parameters: $(\gamma,D)=(2,0.3)$.}
    \label{fig:stationarybump}
\end{figure}

Thus, given parameters $(\beta, \gamma, \theta)$ the full solution to the system can be found by first solving \eqref{eq:thetadelta} numerically for $\Delta$ using a root finding algorithm. This produces at most two valid roots which we denote $0<\Delta_-<\Delta_+<\frac{\pi}{2}$. Then the amplitude of $Q$ in the active region is computed from \eqref{eq:c0} which then allows for $A_0$ to be computed from \eqref{eq:A0}. Finally, $(U(x),Q(x),A(x))$ are found as
\begin{align}
    U(x) = 2c_0\sin(\Delta) \cos(x),\quad Q(x)=\begin{cases}
        c_0,& |x|<\Delta\\
        1,& |x|>\Delta
    \end{cases},\quad A(x)=A_0.\label{eq:stationarybumpsol}
\end{align}
Examples of stationary bump solutions can be seen in Figure \ref{fig:stationarybump}.

\section{Stability of stationary bumps}
\label{sec:stability}

\subsection{Linearization and perturbation ansatz}
\label{sec:linpert}
We examine the stability of bumps using a perturbative analysis.
Assuming parameters lie in a regime where the stationary bump solutions of Section \ref{sec:bumpexistence} exist, we introduce small perturbations to the stationary solutions as follows
\begin{align}
   u(x,t) = U(x)+\epsilon\psi(x,t),\quad q(x,t)= Q(x)+\epsilon\varphi(x,t),\quad a(x,t)= A(x)+\epsilon\eta(x,t).\label{eq:perturbationformula}
\end{align}
with $\epsilon\ll 1$.
The perturbation shifts the bump edges which determine the boundaries of the active region as defined by the threshold condition
\begin{align*}
    u(\pm\Delta+\epsilon\alpha_\pm(t),t)=\theta,
\end{align*}
where $\epsilon \alpha_\pm(t)$ denotes the perturbation to the bump boundary at $x=\pm \Delta$ respectively. A Taylor expansion about $x=\pm \Delta $ yields up to $O(\epsilon)$~\cite{Amari},
\begin{align}
    \alpha_\pm(t) = \pm \frac{\psi(\pm \Delta,t)}{|U^\prime(\Delta)|}:=\pm\mu \psi(\pm \Delta,t).\label{eq:perturbationboundary}
\end{align}
Plugging \eqref{eq:perturbationformula} into \eqref{eq:anfmodel} and dividing through by $\epsilon$ yields evolution equations for the perturbations, which are piecewise-smooth given the step nonlinearity $f(u) = H(u-\theta)$,
\begin{align}
    \psi_t+\psi=&\int_{-\pi}^\pi \cos(x-y) \varphi H(U+\epsilon \psi-\theta)\ dy\label{eq:psi}\\
    &\hspace{0.5in}+\frac{1}{\epsilon}\int_{-\pi}^\pi \cos(x-y) Q\left[H(U+\epsilon \psi -\theta)-H(U-\theta)\right]\ dy,\nonumber\\
    \varphi_t =&\frac{\beta}{\epsilon}Q\left[H(U-\theta)-H(U+\epsilon \psi-\theta)\right]-\gamma A\varphi+\gamma \eta(1-Q)-\beta \varphi H(U+\epsilon\psi-\theta),\label{eq:phi}\\
    \eta_t=&-\varphi_t+D\eta_{xx}.\label{eq:eta}
\end{align}
Taking $\epsilon \to 0$ in \eqref{eq:psi} and carefully evaluating the boundary contributions introduces the functions
\begin{align*}
    Q_+(\psi(\Delta,t)):=\begin{cases}
        1, & \psi(\Delta,t)>0\\
        c_0, & \psi(\Delta,t)< 0
    \end{cases},\quad Q_-(\psi(-\Delta,t)):=\begin{cases}
        1, &\psi(-\Delta,t)>0\\
        c_0, & \psi(-\Delta,t)<0
    \end{cases},
\end{align*}
which depend on $\operatorname{sgn}(\psi(\pm\Delta, t))$.
For the linearized system to yield a spectral problem with time-independent coefficients, $Q_\pm$ must not depend on $t$, which requires that $\psi(\pm\Delta,t)$ does not change sign. This is guaranteed by the exponential ansatz
\begin{align}
    (\psi,\varphi,\eta)(x,t)=e^{\lambda t}(\psi,\varphi,\eta)(x)\label{eq:ansatz}
\end{align}
with $\lambda\in \mathbb{R}$. This restriction excludes complex eigenvalues from the analysis; numerical simulations of the full system \eqref{eq:anfmodel} confirm that the instabilities we observe are real-eigenvalue crossings. This yields
\begin{align}
    (\lambda +1) \psi(x) = \int_{-\Delta^{\sigma_-}}^{\Delta^{\sigma_+}}\cos(x-y) \varphi(y) \ dy +\mu \sum_{s=\pm} Q_s \, \psi(s\Delta) \cos(x - s\Delta), \label{eq:psilinearized}
\end{align}
where $\sum_{s \in \pm}$ denotes summation over $s \in \{ + 1, -1 \}$ and we define $\sigma_{\pm}:=\text{sgn}\left(\psi(\pm \Delta)\right)$ and the endpoints 
\begin{align*}
    \Delta^{\sigma_+}:=\begin{cases}
        \Delta^+, & \sigma_+>0\\
        \Delta^-, & \sigma_+<0
    \end{cases}, \quad \Delta^{\sigma_-}:=\begin{cases}
        -\Delta^+, & \sigma_-<0\\
        -\Delta^-, & \sigma_->0
    \end{cases},
\end{align*}
 indicate the case dependent contributions of $\varphi$ at the stationary bump boundaries $x=\pm \Delta$.

\begin{figure}[t!]
    \centering
    \includegraphics[width=0.95\linewidth]{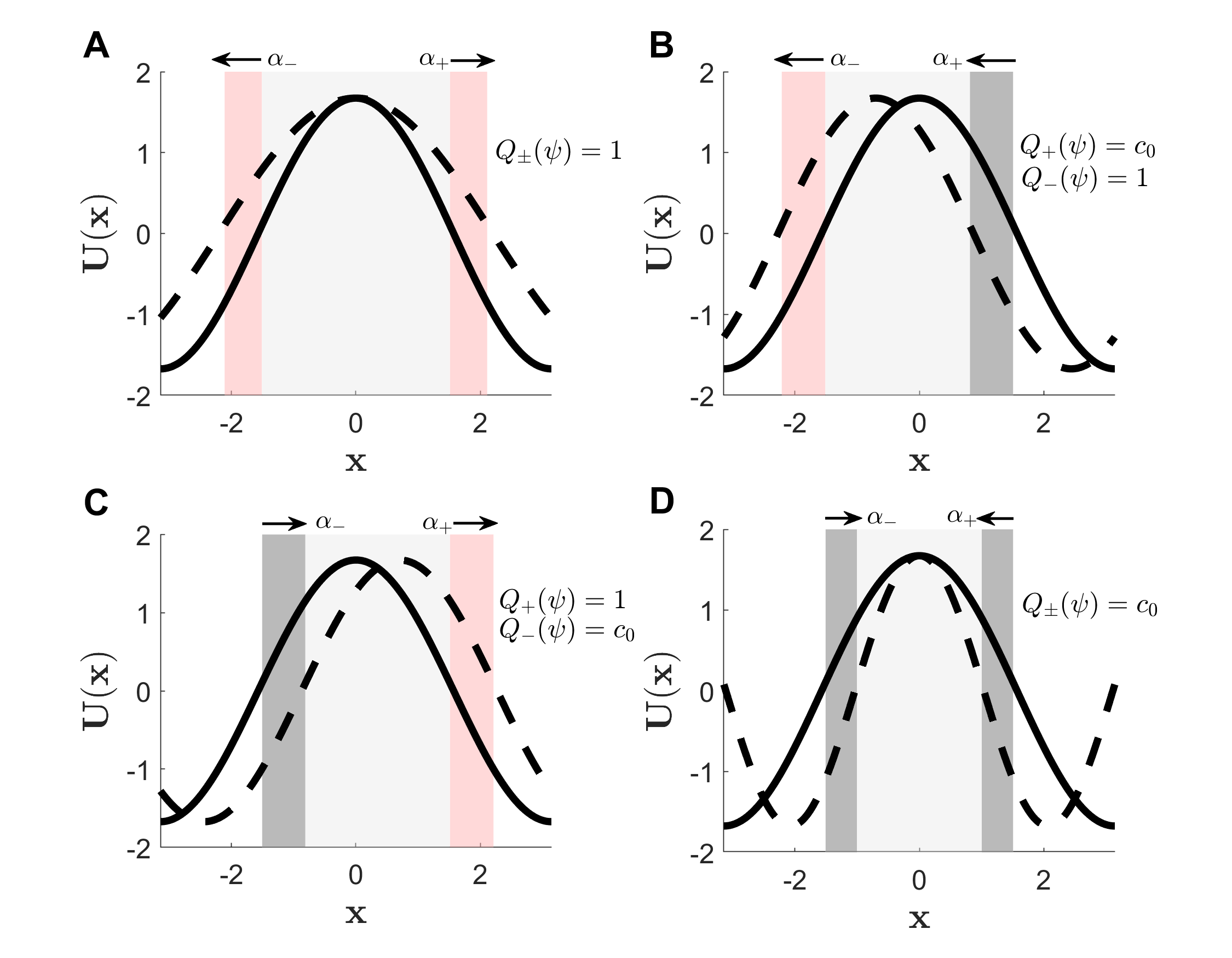}
    \vspace{-2mm}
    \caption{\textbf{Schematic of the four perturbation cases for a stationary bump.} Solid line: stationary bump $U(x)$; dashed line: perturbed profile $U(x) + \epsilon \psi(x,0)$. Light grey shading indicates the active region of the stationary bump. Dark grey shading shows contraction, and red shading shows expansion of the active region in the perturbed profile. {\bf A.} Expansion ($\sigma_\pm = 1$). {\bf B.} Left shift ($\sigma_+ =-1$, $\sigma_-=1$). {\bf C.} Right shift ($\sigma_+=1$, $\sigma_- = -1$). {\bf D.} Contraction ($\sigma_\pm=-1$).}
    \label{fig:perturbationschematic}
\end{figure}
 
 Equation \eqref{eq:perturbationboundary} demonstrates that the sign of the perturbation $\psi$ at the bump boundary determines whether the active region expands or contracts at each end, defining coefficients in the linearized system. Consequently, the stability problem is piecewise-smooth extension of that in \cite{Amari} with four distinct linear stability problems for which $(\sigma_+, \sigma_-) \in \{(\pm 1, \pm 1); (\mp 1, \pm 1) \}$ as indicated in Fig.~\ref{fig:perturbationschematic}. 
 This phenomenon is not unique to our system and has been observed in bump stability analyses of other neural field models with nonlinear negative feedback~\cite{coombes2005bumps,Kilpatrick10b}. In our setting, this piecewise-smooth structure cannot be sidestepped by a judicious choice of the value of $H(0)$, rather the full piecewise-smooth case structure must be embraced~\cite{bressloff2011two,Cihak2024}. The key distinction from prior work is that the astrocytic variable $a$ introduces a spatially extended diffusive field that couples back to the bump boundaries through replenishment, so the piecewise-smooth boundary analysis must be carried alongside the solution of a global elliptic problem for $\eta$.

The assumption that $\lambda \in \mathbb{R}$ is critical: oscillatory perturbations would cause $\psi(\pm\Delta,t)$ to change sign, in contradiction to our framing. As a consequence, our spectral analysis cannot assess the stability of bumps to oscillatory perturbations. However, numerical simulations of the full system \eqref{eq:anfmodel} do not reveal oscillatory instabilities, confirming that the dominant instability mechanism is a real-eigenvalue crossing associated with drift. We also note that the essential spectrum of the linearized operator, determined by the decay rate of the activity equation outside the active region, lies at $\lambda = -1$ and corresponds to perturbation modes that do not affect the bump boundaries~\cite{guo2005existence,Kilpatrick11}. Since these modes decay exponentially, they do not influence bump stability and our analysis focuses on the discrete eigenvalues that govern the dynamics at the bump edges.

\subsection{Projection onto Fourier modes}
Using appropriate trigonometric identities, the right-hand side of \eqref{eq:psilinearized} lies entirely in $\text{span}\{\cos(x),\sin(x)\}$ due to the filtering of neural activity through these modes. Thus, for any eigenvalues $\lambda\neq -1$ we have
\begin{align*}
    \psi(x) = \Psi^c\cos(x)+\Psi^s\sin(x),
\end{align*}
where
\begin{align*}
    \Psi^c:=\frac{1}{\pi}\int_{-\pi}^\pi \psi(x) \cos(x) \ dx,\quad  \Psi^s:=\frac{1}{\pi}\int_{-\pi}^\pi \psi(x) \sin(x) \ dx.
\end{align*}
Substituting this representation into \eqref{eq:psilinearized} and projecting onto $\text{span}\{\cos,\sin\}$ we obtain
\begin{align}
    (\lambda +1) \begin{pmatrix}
        \Psi^c\\ \Psi^s
    \end{pmatrix}= \begin{pmatrix}
        C_\varphi\\ S_\varphi
    \end{pmatrix}+\mathbf{A}(\mathbf{Q})\begin{pmatrix}
        \Psi^c\\\Psi^s
    \end{pmatrix}\label{eq:dispersion1}
\end{align}
where $\mathbf{Q} = (Q_+, Q_-)$ and
\begin{align*}
    \mathbf{A}(\mathbf{Q}) = \mu \begin{pmatrix} 
    (Q_++Q_-)\cos^2(\Delta) & (Q_+-Q_-)\cos(\Delta) \sin(\Delta)\\
    (Q_+-Q_-)\cos(\Delta)\sin(\Delta) & (Q_++Q_-)\sin^{2}(\Delta)
    \end{pmatrix},
\end{align*}
and
\begin{align*}
    C_\varphi = \int_{-\Delta^{\sigma_-}}^{\Delta^{\sigma_+}}\cos(y)\varphi(y) dy,\quad S_\varphi = \int_{-\Delta^{\sigma_-}}^{\Delta^{\sigma_+}}\sin(y)\varphi(y) dy.
\end{align*}

\subsection{Linearized resource equations}
To solve the eigenvalue problem, expressions are needed for $C_\varphi$ and $S_\varphi$, requiring us to solve the perturbative equations for the resource variables $q$ and $a$, \eqref{eq:phi} and \eqref{eq:eta} respectively. The apparent $O(1/\epsilon)$ terms in \eqref{eq:phi} arise from jump discontinuities at the bump boundaries and require care in our interpretation of how piecewise smoothness bears upon their final form. Consider
\begin{align*}
    \lim_{\epsilon\to 0}\frac{\beta}{\epsilon}&\langle Q(H(U-\theta)-H(U+\epsilon\psi-\theta)), f \rangle =\lim_{\epsilon\to 0}\frac{\beta}{\epsilon}\left[\int\limits_{-\Delta}^{-\Delta+\epsilon\alpha_-(t)}\!\!\!\!\!\!\!\! Q(x)f(x) \ dx-\!\!\!\!\!\!\!\int\limits_{\Delta}^{\Delta+\epsilon \alpha_+(t)}\!\!\!\!\!\!\!Q(x)f(x)\ dx\right]\\
    &= -\beta\mu \lim_{\epsilon\to 0}\left[(\psi(\Delta,t)Q(\Delta+\epsilon\alpha_+(t)) f(\Delta) -\psi(-\Delta,t) Q(-\Delta+\epsilon\alpha_-(t))f(-\Delta))\right]\\
    &=-\mu\beta\langle  \psi(\Delta)Q_+\delta(x-\Delta)+Q_-\psi(-\Delta)\delta(x+\Delta),f\rangle
\end{align*}
where we note
\begin{align*}
    \lim_{\epsilon\to 0}Q(\Delta+\epsilon\alpha_+(t))= Q_+,\quad \lim_{\epsilon\to 0}Q(-\Delta+\epsilon\alpha_-(t))= Q_-.
\end{align*}
So, as $\epsilon\to 0$
\begin{align*}
    \frac{\beta}{\epsilon}Q\left[H(U-\theta)-H(U+\epsilon \psi-\theta)\right]=-\mu\beta(\psi(\Delta) Q_+\delta(x-\Delta)+\psi(-\Delta)Q_-\delta(x+\Delta)) 
\end{align*}
in the sense of distributions. This allows us to write the perturbed resource system in linearized form as
 \begin{align}
     \lambda \varphi &= -\mu\beta \left(\psi(\Delta)Q_+\delta(x-\Delta)+\psi(-\Delta)Q_-\delta(x+\Delta)\right) -\gamma A_0\varphi +\gamma \eta (1-Q)-\beta \chi^{\sigma} \varphi\label{eq:philinearized}\\
      \lambda \eta &= -\lambda \varphi +D\eta_{xx}\label{eq:etalinearized}
 \end{align}
 where $\chi^{\sigma}(x)$ is an indicator function over the active region of the perturbed bump, whose support depends on the particular case of $(\sigma_+, \sigma_-)$ as described above.

 \subsection{Neutral translation mode}
 As a consistency check, we verify that the linearized system ($\ref{eq:dispersion1}-\ref{eq:etalinearized}$) preserves the neutral mode expected from the translational invariance of the original system.
The infinitesimal translation mode is obtained by spatial differentiation,
 \begin{align*}
    \psi(x)&=U^\prime(x)=-2c_0\sin(\Delta)\sin(x),\\
    \varphi(x)&=Q^\prime(x) = (1-c_0)\left[\delta(x-\Delta)-\delta(x+\Delta)\right],\\
    \eta(x)&=A^\prime(x)=0.
 \end{align*}
 For $\lambda=0$ it is clear that \eqref{eq:etalinearized} is satisfied by $\eta=0$. Examining $\psi$ and $\varphi$ we note that $\psi(\Delta)<0$ and $\psi(-\Delta)>0$, which implies that $Q_+=c_0$, $Q_-=1$ and $-\Delta^{\sigma_-}<-\Delta<\Delta^{\sigma_+}<\Delta$. Thus,
 \begin{align*}
     &\int_{-\Delta^{\sigma_-}}^{\Delta^{\sigma_+}}\cos(y) \varphi(y) \ dy = -(1-c_0)\cos(\Delta),\quad \int_{-\Delta^{\sigma_-}}^{\Delta^{\sigma_+}}\sin(y) \varphi(y) \ dy = (1-c_0)\sin(\Delta).
 \end{align*}
 So, using the definition of $\mu$ we have
 \begin{align*}
     \psi(x) =& (1-c_0)\sin(\Delta)\sin(x) -(1-c_0)\cos(\Delta)\cos(x) \\
     &-c_0\left[\cos(\Delta)\cos(x)+\sin(\Delta)\sin(x)\right]+\left[\cos(\Delta)\cos(x)-\sin(\Delta)\sin(x)\right]\\
     =&-2c_0\sin(\Delta)\sin(x),
 \end{align*}
 and so $\psi(x)=U^\prime(x)$. For $\varphi$, letting $\mathbf{1}_{\mathcal A}(x)$ denote the indicator function of a set ${\mathcal A}$ and setting $\eta(x)=0$ and $\psi(x)=U^\prime(x)$ we have
 \begin{align*}
     \left[\gamma A_0+\beta \mathbf{1}_{-\Delta\leq x<\Delta}\right]\varphi=\mu\beta \left(2c_0^2\sin^2(\Delta)\delta(x-\Delta)-2c_0\sin^2(\Delta)\delta(x+\Delta)\right)
 \end{align*}
 which, noting that $\mu=|U^\prime(\Delta)|^{-1}$ and that $c_0=(\beta+\gamma A_0)^{-1}(\gamma A_0)$, implies 
 \begin{align*}
     \varphi = (1-c_0)\left[\delta(x-\Delta)-\delta(x+\Delta)\right].
 \end{align*}
Thus, the expected equalities for the neutral translation mode hold upon selecting for the correct linearized system.

\subsection{Construction of the Evans function}
Returning to the eigenvalue problem, from \eqref{eq:philinearized}, $\varphi$ can be solved for as
\begin{align*}
    \varphi &= \frac{1}{\lambda+\gamma A_0+\beta\chi^{\sigma}(x)}\left[\gamma \eta(1-c_0)\mathbf{1}_{|x|<\Delta}-\mu\beta \left(\psi(\Delta)Q_+\delta(x-\Delta)+\psi(-\Delta)Q_-\delta(x+\Delta)\right)\right].
\end{align*}
Then substituting the above expression into \eqref{eq:etalinearized} yields an ordinary differential equation for $\eta$
\begin{align*}
    \mathcal{L}_\lambda\eta = Q_+\psi(\Delta) q(x)\delta(x-\Delta)+Q_-\psi(-\Delta) q(x)\delta(x+\Delta)
\end{align*}
where $\mathcal{L}_\lambda:=(p(x)-\partial_{xx})$ and
\begin{align*}
    p(x) &= \frac{\lambda(\lambda+\gamma A_0+\beta\chi^{\sigma}(x)+\gamma (1-c_0)\mathbf{1}_{|x|<\Delta})}{D(\lambda+\gamma A_0+\beta\chi^{\sigma}(x))},\quad
    q(x) = \frac{\lambda\mu\beta}{D(\lambda+\gamma A_0+\beta\chi^{\sigma}(x))},
\end{align*}
Assuming that $\lambda \neq -\gamma A_0-\beta$ and $\lambda\neq -\gamma A_0$ then the equation $\mathcal{L}_\lambda G_\lambda = \delta(x-y)$ is solvable, implying that $\mathcal{L}_\lambda$ is invertible.
Defining $g_\pm=\mathcal{L}_\lambda^{-1}(q(x)\delta(x\mp \Delta))$ gives $\eta = Q_+\psi(\Delta) g_++Q_-\psi(-\Delta)g_-$ and so it suffices to solve
\begin{align}
    \label{eq:g}\mathcal{L}_\lambda g_\pm =q(x)\delta(x\mp \Delta),\quad \text{ on } [-\pi,\pi],
\end{align}
with $2\pi$-periodic boundary conditions.
The functions $g_{\pm}$ describe how a localized perturbation at each bump boundary propagates through the astrocyte layer via diffusion. As we will show, this spatial smoothing of resource inhomogeneities contributes to the stabilization of stationary bumps.
The source points in \eqref{eq:g} split the equation into three distinct regions where the dynamics reduce to a constant-coefficient ODE, specifically $R_1=(-\pi,-\Delta)$, $R_2=(-\Delta,\Delta)$ and $R_3=(\Delta,\pi)$. On $R_i$ one finds $g_\pm(x)=k_ie^{\sqrt{c_i}x}+\kappa_ie^{-\sqrt{c_i}x}$ where 
\begin{align*}
    c_1=c_3=\frac{\lambda}{D},\quad c_2=\frac{ \lambda(\lambda+\gamma A_0+\beta +\gamma(1-c_0))}{D(\lambda+\gamma A_0+\beta)},
\end{align*}
and the coefficients $k^\pm_i$ and $\kappa^\pm_i$ are found from suitable conditions. Specifically, we enforce periodicity of $g_\pm$, continuity at $x=\pm \Delta$, continuity of the derivative across the non-singular interface and jump conditions:
\begin{align*}
    [g_+^\prime]_\Delta&=-\frac{\lambda\mu\beta}{D(\lambda+\gamma A_0+\beta\chi^{\sigma}(\Delta))},\quad
    [g_-^\prime]_{-\Delta}=-\frac{\lambda\mu\beta}{D(\lambda+\gamma A_0+\beta\chi^{\sigma}(-\Delta))}
\end{align*}
at the source point. Having found $g_\pm$,
$(C_\varphi, S_\varphi)$ can be rewritten in terms of $\eta$ and solved as
\begin{align*}
    \begin{pmatrix}
        C_\varphi\\ S_\varphi
    \end{pmatrix} &=\mathbf{B}(\lambda, \mathbf{Q})\begin{pmatrix} \Psi^c\\ \Psi^s\end{pmatrix}+\mathbf{C}(\lambda, \mathbf{Q})\begin{pmatrix}
        \Psi^c\\ \Psi^s
    \end{pmatrix}
\end{align*}
where
\begin{align*}
    B_{11}&=\frac{\gamma(1-c_0)}{\lambda+\beta+\gamma A_0}\int_{-\Delta}^\Delta\cos(y)\cos(\Delta)\left[Q_+g_+(y)+Q_-g_-(y)\right]dy \\
    B_{12}&=\frac{\gamma(1-c_0)}{\lambda+\beta+\gamma A_0}\int_{-\Delta}^\Delta\cos(y)\sin(\Delta)\left[Q_+g_+(y)+Q_-g_-(y)\right]dy,\\
    C_{11}&=-\mu\beta\int_{-\Delta^{\sigma_-}}^{\Delta^{\sigma_+}}\frac{\cos(y)\cos(\Delta)}{\lambda +\gamma A_0+\beta\chi^{\sigma}(y)}\left[Q_+\delta(y-\Delta)+Q_-\delta(y+\Delta)\right]\ dy \\
    C_{12}&=-\mu\beta\int_{-\Delta^{\sigma_-}}^{\Delta^{\sigma_+}}\frac{\cos(y)\sin(\Delta)}{\lambda +\gamma A_0+\beta\chi^{\sigma}(y)}\left[Q_+\delta(y-\Delta)-Q_-\delta(y+\Delta)\right]dy,
\end{align*}
with analogous expressions for $B_{21}, B_{22}, C_{21}$ and $C_{22}$ obtained by replacing $\cos(x)$ with $\sin(x)$.
Returning to \eqref{eq:dispersion1} one obtains
\begin{align}
    (\lambda +1) \begin{pmatrix}
        \Psi^c\\ \Psi^s
    \end{pmatrix}= \bigg[\mathbf{A}(\mathbf{Q})+\mathbf{B}(\lambda,\mathbf{Q})+\mathbf{C}(\lambda,\mathbf{Q})\bigg]\begin{pmatrix}
        \Psi^c\\\Psi^s
    \end{pmatrix}.\label{eq:dispersion2}
\end{align}
Here $\mathbf{A}$ encodes the direct effect of bump boundary shifts on neural activity, $\mathbf{B}$ represents the indirect contribution of astrocytic diffusion through the resource coupling, and $\mathbf{C}$ captures the immediate synaptic resource response at the bump edges.
Thus, the stability problem reduces to a nonlinear eigenvalue problem, where the roots of
 \begin{align}
     \mathcal{E}(\lambda)=\text{det}\left[(\lambda+1)\mathbf{I}-\mathbf{C}(\lambda,\mathbf{Q})-\mathbf{B}(\lambda, \mathbf{Q})-\mathbf{A}(\mathbf{Q})\right].\label{eq:determinant}
 \end{align}
determine the stability of the system. We refer to $\mathcal{E}(\lambda)$ as the Evans function of the linearized system~\cite{coombes2004evans}. Its roots, which determine the stability of stationary bumps, must be found numerically due to the transcendental dependence on $\lambda$ in $\mathbf{B}$ and $\mathbf{C}$.

 \subsection{Limiting Cases}
 \label{sec:limits}
 In contrast to previous models, \eqref{eq:anfmodel} contains three parameters relevant to resource redistribution. To better understand the role of astrocytic diffusion $D$ and synaptic replenishment $\gamma$ we examine two limiting cases: when $D=0$ and when $D\to\infty$.
 
 \subsubsection{Strong diffusion limit $D \to \infty$} In the strong diffusion limit the form of the stationary solutions is unchanged. Thus, starting with \eqref{eq:eta}, dividing through by $D$ and taking $D\to \infty$ forces $\eta_{xx}=0$. Periodicity then implies that $\eta(x)=\eta_0$ is constant. The conservation constraint $\int_{-\pi}^{\pi} \left[ \varphi(x) + \eta (x) \right] dx = 0 $ then ensures
 \begin{align*}
     \eta_0=-\overline{\varphi}:=-\frac{1}{2\pi}\int_{-\pi}^\pi \varphi(x) \ dx. 
 \end{align*}
 As a result, \eqref{eq:phi} becomes
\begin{align*}
   \left[\lambda+\gamma A_0+\beta \chi^{\sigma}(x)\right]\varphi(x)+\gamma (1-c_0)\mathbf{1}_{|x|<\Delta}\overline{\varphi} &= -\mu\beta \sum_{s = \pm} \psi(s \Delta) Q_s + \delta(x-s \Delta).
\end{align*}
Provided $\lambda+\gamma A_0+\beta\chi^\sigma(\pm\Delta)\neq 0$, we obtain
\begin{align}
    \varphi&=-\mu\beta \sum_{s=\pm}\frac{\psi(s\Delta)Q_s}{\lambda+\gamma A_0+\beta\chi^{\sigma}(s\Delta)}\delta(x-s\Delta)- \overline{\varphi}\frac{\gamma(1-c_0)\mathbf{1}_{|x|<\Delta}}{\left[\lambda+\gamma A_0+\beta\chi^{\sigma}(x)\right]}. \label{eq:Dinfphi}
\end{align}
The average of $\varphi$ can be found by integrating \eqref{eq:Dinfphi} over $(-\pi,\pi)$ and isolating $\overline{\varphi}$ as
\begin{align*}
    \overline{\varphi}=-\frac{\mu\beta(\lambda+\gamma A_0+\beta)}{2\pi(\lambda+\gamma A_0+\beta)+2\Delta\gamma(1-c_0)}\left[\frac{\psi(\Delta)Q_+}{ \lambda+\gamma A_0+\beta \chi^{\sigma}(\Delta)}+\frac{\psi(-\Delta)Q_-}{\lambda+\gamma A_0+\beta\chi^{\sigma}(-\Delta)}\right]
\end{align*}
Returning to $C_\varphi$ and $S_\varphi$ we have
\begin{align*}
 \begin{pmatrix}
         C_\varphi\\
         S_\varphi
     \end{pmatrix}=\mathbf{B}_\infty(\lambda, \mathbf{Q})\begin{pmatrix}
         \Psi^c\\\Psi^s
     \end{pmatrix}
 \end{align*}
where the entries of $\mathbf{B}_\infty(\lambda,\mathbf{Q})$ are obtained by 
integrating \eqref{eq:Dinfphi} against $\cos(x)$ or $\sin(x)$ and expanding 
$\psi(\pm\Delta)=\Psi^c\cos\Delta\pm\Psi^s\sin\Delta$. Substituting into 
\eqref{eq:dispersion1}, bump stability in the strong diffusion limit is 
determined by the zeros of
\begin{align}
    \mathcal{E}_\infty(\lambda) = \det\left[(\lambda+1)\mathbf{I} 
    - \mathbf{B}_\infty(\lambda,\mathbf{Q}) 
    - \mathbf{A}(\mathbf{Q})\right], \label{eq:evansinf}
\end{align}
whose roots must be found numerically. By translational symmetry $\lambda=0$ 
is always a root, and the stability boundary in parameter space is again 
characterized by the double-root condition 
$\mathcal{E}_\infty(0)=\mathcal{E}_\infty'(0)=0$, computed numerically via 
a centered difference approximation to $\mathcal{E}_\infty'$.

\subsubsection{Zero diffusion limit $D=0$}
When $D=0$, the solution in Section~\ref{sec:bumpexistence} no longer holds as $A(x)$ becomes piecewise constant. However, stationary bumps can still be shown to exist: in the absence of diffusion, resource balance holds locally so that $q(x,t)+a(x,t)=1$, and the fields evolve as
\begin{align}
    \frac{\partial u}{\partial t}(x,t) &= -u(x,t) + \int_{-\pi}^{\pi} \cos(x-y)\,q(y,t)\,H(u(y,t)-\theta)\,dy, \label{eq:redmodel}\\
    \frac{\partial q}{\partial t}(x,t) &= \gamma(1-q(x,t))^2 - \beta H(u(x,t)-\theta)\,q(x,t).\nonumber
\end{align}
The stationary bump equations then reduce to
\begin{align*}
    U(x) = \int_{-\Delta}^{\Delta}\cos(x-y)\,Q(y)\,dy, \qquad \gamma(1-Q(x))^2 = \beta\bigl[H(x+\Delta)-H(x-\Delta)\bigr] Q(x),
\end{align*}
with solution
\begin{align}
    U(x) = 2c_0\sin(\Delta)\cos(x), \quad Q(x) = \begin{cases} c_0, & |x|<\Delta\\ 
    1, & |x|>\Delta \end{cases}, \qquad c_0 := \frac{(2\gamma+\beta)-\sqrt{\beta^2+4\beta\gamma}}{2\gamma}, 
    \label{eq:stationarybumpsolD=0}
\end{align}
with half-width $\Delta$ again determined by the self-consistency condition 
$\theta = c_0\sin(2\Delta)$, subject to $2\gamma+\beta>\sqrt{\beta^2+4\beta\gamma}$, 
with $0<\Delta_-<\Delta_+<\pi/2$.

Clearly $D=0$ is a singular limit since $A(x)=1-Q(x)\neq\text{const}$: the order of the spatial operator drops, changing the function space for eigenfunctions, so the Evans function \eqref{eq:determinant} cannot be analytically continued to $D=0$. 
We therefore derive stability independently for this limit. Linearizing \eqref{eq:redmodel} with $(u,q)=(U,Q)+\epsilon e^{\lambda t}(\psi(x),\varphi(x))$, $\lambda\in\mathbb{R}$, and following the same boundary perturbation analysis as in Section~\ref{sec:linpert} gives
\begin{align}
    (\lambda+1)\psi &= \int_{-\Delta^{\sigma_-}}^{\Delta^{\sigma_+}}\cos(x-y)\,\varphi(y)\,dy + \mu\sum_{s=\pm 1}Q_s\,\psi(s\Delta)\cos(x-s\Delta), \label{eq:psi_lin}\\
    \lambda\,\varphi &= -\bigl[2\gamma(1-Q(x))+\beta\chi^\sigma(x)\bigr]\varphi - \beta\mu\sum_{s=\pm 1}Q_s\,\psi(s\Delta)\,\delta(x-s\Delta). \label{eq:phi_lin}
\end{align}
Since \eqref{eq:phi_lin} is purely local, $\varphi$ vanishes in the interior of both regions if we assume that $\lambda\neq - \left[ \gamma (1-c_0) + \beta \chi^{\sigma}( \pm \Delta) \right]$. Then, $\varphi$ is supported entirely at the bump boundaries:
\begin{align}
    \varphi(x) = -\beta\mu\sum_{s=\pm 1}\frac{Q_s\,\psi(s\Delta)}
    {\lambda+\mu_s}\,\delta(x-s\Delta), \label{eq:phi_solved}
\end{align}
where $\mu_s := \gamma(1-c_0)+\beta\chi^\sigma(s\Delta)$. Writing $\varphi(x) = \varphi_+\delta(x-\Delta)+\varphi_-\delta(x+\Delta)$, the coefficients are given explicitly by
\begin{align*}
    \varphi_s = -\frac{\beta\mu Q_s\,\psi(s\Delta)}{\lambda+\mu_s}, \quad s=\pm 1.
\end{align*}
Substituting \eqref{eq:phi_solved} into \eqref{eq:psi_lin}, the 
$\delta$-mass at $s\Delta$ contributes to the integral only if 
$\sigma_s>0$, i.e.\ only if that edge lies within the perturbed 
active region. Defining
\begin{align*}
    \Xi_s(\lambda) := \begin{cases} \dfrac{\lambda+\mu_s-\beta Q_s}{\lambda+\mu_s}, & \sigma_s > 0 \\[6pt] 1, & \sigma_s < 0 \end{cases}
\end{align*}
to encode whether the $\delta$-mass at $s\Delta$ is included or 
excluded, the reduced equation is
\begin{align}
    (\lambda+1)\psi(x) = \mu \sum_{s=\pm 1}Q_s\,\psi(s\Delta)\,
    \Xi_s(\lambda)\cos(x-s\Delta). \label{eq:reduced}
\end{align}
Evaluating \eqref{eq:reduced} at $x=\pm\Delta$ gives a $2\times 2$ nonlinear eigenvalue problem in $(\psi(\Delta),\psi(-\Delta))$, since $\Xi_s(\lambda)$ depends on $\lambda$ through $\mu_s$. Setting the determinant to zero and clearing the denominators $(\lambda+\mu_s)$ yields a polynomial characteristic equation whose degree depends on the case. For translation, this is a cubic which factors as $\lambda\cdot\tilde{\mathcal{E}}_{\rm tr}(\lambda)=0$ due to marginal stability of shifts, leaving a quadratic for the remaining roots. For expansion, the matrix is proportional to a constant matrix and the determinant factors into two independent quadratics. For contraction, $\Xi_s$ reduces to unity and the characteristic equations are linear. This case-dependence of the degree of the characteristic equation is a general feature of piecewise-smooth stability problems~\cite{bressloff2011two,bernardo2008piecewise}, reflecting the fact that different sign patterns activate different numbers of boundary contributions and thus change the effective dimension of the spectral problem.

\paragraph{Translation} ($\sigma_+>0,\,\sigma_-<0$, so $Q_+=1$, $Q_-=c_0$). 
The right edge expands and the left contracts, giving the asymmetric system
\begin{align}
    \begin{pmatrix}(\lambda+1)-\mu \Xi_+(\lambda) & -\mu c_0\cos 2\Delta\\ 
    -\mu \Xi_+(\lambda)\cos 2\Delta & (\lambda+1)-\mu c_0\end{pmatrix}
    \begin{pmatrix}\psi(\Delta)\\\psi(-\Delta)\end{pmatrix} = 0,
    \label{eq:2x2_tr}
\end{align}
where $\Xi_+(\lambda) = (\lambda+\gamma(1-c_0))/(\lambda+\gamma(1-c_0)+\beta)$.
Setting the determinant to zero and clearing the denominator 
$(\lambda+\gamma(1-c_0)+\beta)$ yields a cubic in $\lambda$:
\begin{align}
    \mathcal{E}_{\rm tr}(\lambda) :=\, \lambda^3 + b_1\lambda^2 + b_0\lambda = 0,
    \label{eq:char_tr}
\end{align}
where the constant term vanishes since $\lambda=0$ is an exact root by 
translation symmetry, verified using $\gamma(1-c_0)^2=\beta c_0$ and simplifying. 
Factoring out $\lambda$ gives the reduced quadratic
\begin{align}
    \tilde{\mathcal{E}}_{\rm tr}(\lambda) := \lambda^2 + b_1\lambda + b_0 = 0,
    \label{eq:char_tr_reduced}
\end{align}
where
\begin{align*}
    b_1 = 2+\frac{\gamma(1-c_0)}{c_0}-\mu(1+c_0), \qquad
    b_0 = \mu(1-c_0)(1-2\gamma)\cos 2\Delta.
\end{align*}
The bump is stable to drift if and only if $b_0>0$ and $b_1>0$. By reflection symmetry, left-shift perturbations yield an identical characteristic equation.
Since $\mu>0$, $(1-c_0)>0$, and $\cos 2\Delta<0$ on the upper branch ($\Delta>\pi/4$), we have $b_0>0$ if and only if $\gamma>1/2$, which is the binding stability condition throughout the parameter ranges we consider.

\paragraph{Expansion} ($\sigma_+=\sigma_->0$, so $Q_+=Q_-=1$). Both edges expand and both $\delta$-masses are included. Since $\psi(\Delta)>0$ and $\psi(-\Delta)>0$, only the symmetric eigenvector $\psi(\Delta)=\psi(-\Delta)$ is consistent, giving the single characteristic equation after clearing $(\lambda+\gamma(1-c_0)+\beta)$:
\begin{align}
    \mathcal{E}_{\rm exp}(\lambda) := \lambda^2 + \left(1+\frac{\gamma(1-c_0)}{c_0}- 2\mu\cos^2\!\Delta\right)\lambda + \frac{\gamma(1-c_0)}{c_0}\left(1 - 2\mu c_0\cos^2\!\Delta\right) = 0.        \label{eq:char_exp}
\end{align}
It can be shown that the Routh--Hurwitz conditions for \eqref{eq:char_exp} are satisfied, so stability is determined by whether $b_0>0$ from the translation case.

\paragraph{Contraction} ($\sigma_+=\sigma_-<0$, so $Q_+=Q_-=c_0$, $\Xi_\pm=1$). Both edges contract and both $\delta$-masses are excluded. Since $\psi(\Delta)<0$ and $\psi(-\Delta)<0$, only the symmetric eigenvector $\psi(\Delta)=\psi(-\Delta)$ is consistent, giving
\begin{align}
    \mathcal{E}_{\rm con}(\lambda) := (\lambda+1) - \mu c_0(1+\cos 2\Delta) = 0, \label{eq:char_con}
\end{align}
with explicit root $\lambda_{\rm con} = \cos 2\Delta/\sin^2\Delta$, which is strictly negative for $\Delta>\pi/4$ (upper branch) and strictly positive for $\Delta<\pi/4$ (lower branch), confirming that lower-branch bumps are unstable to symmetric contraction.
Hence, wide bumps for $D=0$ are linearly stable if and only if $\gamma>1/2$.

\section{Numerical validation and mechanistic interpretation}
\label{sec:numerics}

\begin{figure}[t!]
    \centering
    \includegraphics[width=0.8\linewidth]{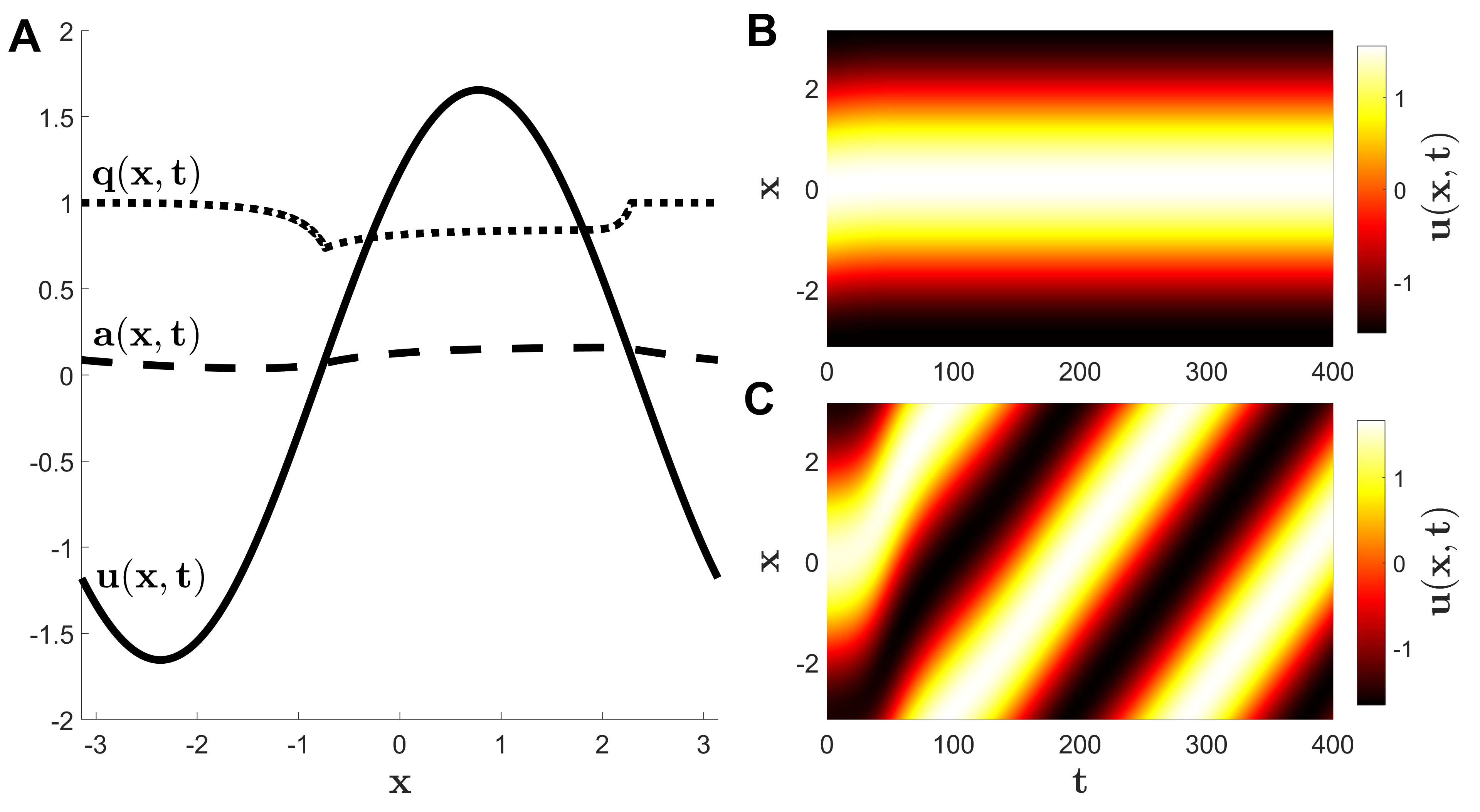}
    \caption{{\bf Numerical simulations of shift perturbations.} Behavior of stationary bump solutions in response to right shift perturbations. Initial conditions are of the form of \eqref{eq:perturbationIC} with parameters $(\beta,\gamma,D,\theta)=(0.06,2,D,0.1)$ for $D$ as specified. Solutions were simulated up to $T=400$. {\bf A.} Cross sections of $u$, $q$, and $a$ for $D=0.05$ at $T=400$. {\bf B.} Time evolution of $u(x,t)$ for $D=0.7$. {\bf C.} Time evolution of $u(x,t)$ for $D=0.05$. Increasing astrocytic diffusion $D$ stabilizes otherwise unstable stationary bumps.}
    \label{fig:travelingwaves}
\end{figure}

\subsection{Stability boundary and numerical implementation}
\label{sec:stabilityboundary}
We focus on the stability of bump solutions under a rightward shift perturbation, 
corresponding to $\sigma_+>0$ and $\sigma_-<0$, so that the right edge expands 
and the left edge contracts, giving
\begin{align*}
    \chi^{\sigma}(x) = \mathbf{1}_{-\Delta < x \leq \Delta},
\end{align*}
where the asymmetry between the open left and closed right endpoints reflects 
the inclusion of the $\delta$-mass at $x=\Delta$ but not at $x=-\Delta$ in 
the perturbed active region. Hence, the integral contributions to 
$\mathcal{E}(\lambda)$ depend only on the interface at $x=\Delta$. Stability 
is determined by the zeros of $\mathcal{E}(\lambda)$, with instability occurring 
when a zero crosses into the positive real axis. By translational symmetry, 
$\lambda=0$ is always a root of \eqref{eq:determinant}. Numerical investigation 
of $\mathcal{E}(\lambda)$ shows that the transition between stable and unstable 
parameter regimes occurs when this root becomes a double root, so the stability 
boundary in parameter space is characterized by
\begin{align}
    \mathcal{E}(0) = 0 \quad \text{and} \quad \mathcal{E}'(0) = 0, 
    \label{eq:transitionboundary}
\end{align}
where the derivative is approximated numerically by a centered difference.

\subsection{Simulation method}
Numerical simulations of \eqref{eq:anfmodel} were carried out on a uniform grid 
with spacing $\Delta x = \pi/4000$ with periodic boundary conditions. Diffusion 
in the astrocytic resource variable was discretized using a second-order centered 
finite difference scheme. The neural activity variable was updated using explicit 
Euler, the astrocyte resource variable using implicit Euler, and the synaptic 
resource variable semi-analytically via an exponential integrating factor. For 
parameters $(\beta,\gamma,D,\theta)$ in a regime where stationary bump solutions 
exist, the stationary solution \eqref{eq:stationarybumpsol} was perturbed as
\begin{align}
    (u(x,t),q(x,t),a(x,t))= (U(x),Q(x),A(x))+\epsilon (\sin(x),0,0)\label{eq:perturbationIC}
\end{align}
where $\epsilon = 0.05\times U(0)$. Under this perturbation, bump solutions 
exhibit two distinct behaviors depending on the parameter regime: either 
continuous drift, or an initial displacement followed by relaxation to a 
stationary bump at a new location (see Figure~\ref{fig:travelingwaves}). 
Bumps were classified as stable if the velocity of the bump's center of mass 
eventually decayed to zero.

\begin{figure}[t!]
    \centering
    \includegraphics[width=\linewidth]{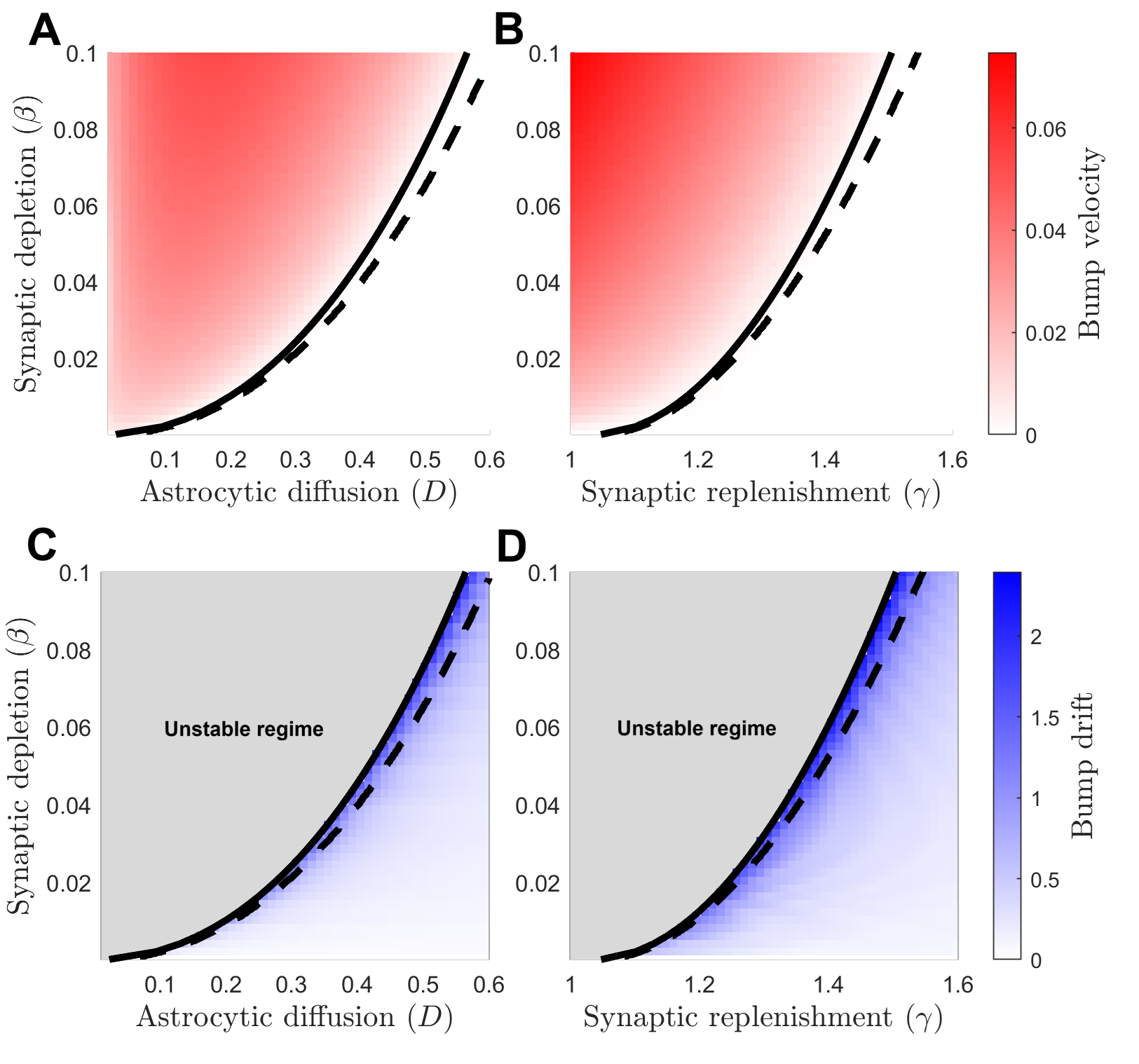}
    \caption{{\bf Phase diagrams of bump velocity and drift.} Bump velocity ({\bf A, B}) and total drift distance ({\bf C, D}) at $T=500$ following the small rightward kick \eqref{eq:perturbationIC}, for simulations of \eqref{eq:anfmodel}. Solid black line: Evans function stability boundary \eqref{eq:transitionboundary}; dashed line: Fourier truncation boundary. Light grey region ({\bf C, D}): predicted unstable regime. {\bf A, C.} $(\beta,\gamma,D,\theta)=(\beta,2,D,0.1)$. {\bf B, D.} $(\beta,\gamma,D,\theta)=(\beta,\gamma,1,0.1)$.}
    \label{fig:phasediagrams}
\end{figure}

\subsection{Comparison with Evans function and Fourier truncation predictions}
The Evans function stability boundary \eqref{eq:transitionboundary} was compared 
against both direct numerical simulations and a low-dimensional Fourier truncation 
approach similar to that of \cite{Kilpatrick10b,Kilpatrick11}. Rather than making the exponential ansatz \eqref{eq:ansatz} and directly 
linearizing \eqref{eq:psi}--\eqref{eq:eta}, the truncation method mollifies the 
evolution equations for $(\varphi,\eta)$ by projecting onto $\{\cos,\sin\}$, 
defining
\begin{align*}
    \Phi^c(t) := \int_{-\Delta^{\sigma_-}}^{\Delta^{\sigma_+}}\varphi(y,t)\cos(y)\,dy, 
    \quad 
    \Phi^s(t) := \int_{-\Delta^{\sigma_-}}^{\Delta^{\sigma_+}}\varphi(y,t)\sin(y)\,dy,
\end{align*}
with analogous definitions $H^c(t)$, $H^s(t)$ for $\eta$. We restrict to 
perturbations of the form
\begin{align*}
    \eta(x,t) = \eta^c(t)\cos(x) + \eta^s(t)\sin(x).
\end{align*}
Substituting into \eqref{eq:phi} and integrating against $\cos(y)$ and 
$\sin(y)$ over the perturbed active region gives
\begin{align}
    \dot{\Phi}^c &= -(\gamma A_0+\beta)\Phi^c +\gamma(1-c_0)H^c \notag\\
    &\quad - \mu\beta\left[(\Theta_++\Theta_-)\Psi^c\cos^2\Delta 
    + (\Theta_+-\Theta_-)\Psi^s\sin\Delta\cos\Delta\right], \label{eq:Phic}\\
    \dot{\Phi}^s &= -(\gamma A_0+\beta)\Phi^s + \gamma(1-c_0)H^s \notag\\
    &\quad - \mu\beta\left[(\Theta_+-\Theta_-)\Psi^c\sin\Delta\cos\Delta 
    + (\Theta_++\Theta_-)\Psi^s\sin^2\Delta\right]. \label{eq:Phis}
\end{align}
where $\Theta_\pm:=H(\psi(\pm\Delta,t)).$
Similarly, substituting into \eqref{eq:etalinearized}, integrating against 
$\cos(y)$ and $\sin(y)$ over $[-\pi,\pi]$, and using $\eta_{xx} = -\eta$, gives
\begin{align}
    \dot{H}^c = -\dot{\Phi}^c - DH^c, \qquad \dot{H}^s = -\dot{\Phi}^s - DH^s. \label{eq:Hs}
\end{align}
Together with the projected $\psi$-equation
\begin{align*}
    \begin{pmatrix}\dot \Psi^c\\\dot \Psi^s\end{pmatrix} = -\begin{pmatrix}
        \Psi^c\\ \Psi^s
    \end{pmatrix}+
    \begin{pmatrix}\Phi^c\\\ \Phi^s\end{pmatrix} + 
    \mathbf{A}(Q_+,Q_-)\begin{pmatrix}\Psi^c\\\Psi^s\end{pmatrix},
\end{align*}
equations \eqref{eq:Phic}--\eqref{eq:Hs} form a closed $6\times 6$ linear ODE 
system in $(\Psi^c,\Psi^s,\Phi^c,\Phi^s,H^c,H^s)$. Seeking solutions 
proportional to $e^{\lambda t}$ reduces this to a linear algebraic system whose 
characteristic polynomial approximates the Evans function. Stability of the bump 
is then assessed by computing the eigenvalues of the resulting matrix numerically.

\begin{figure}[t!]
    \centering
    \includegraphics[width=0.8\linewidth]{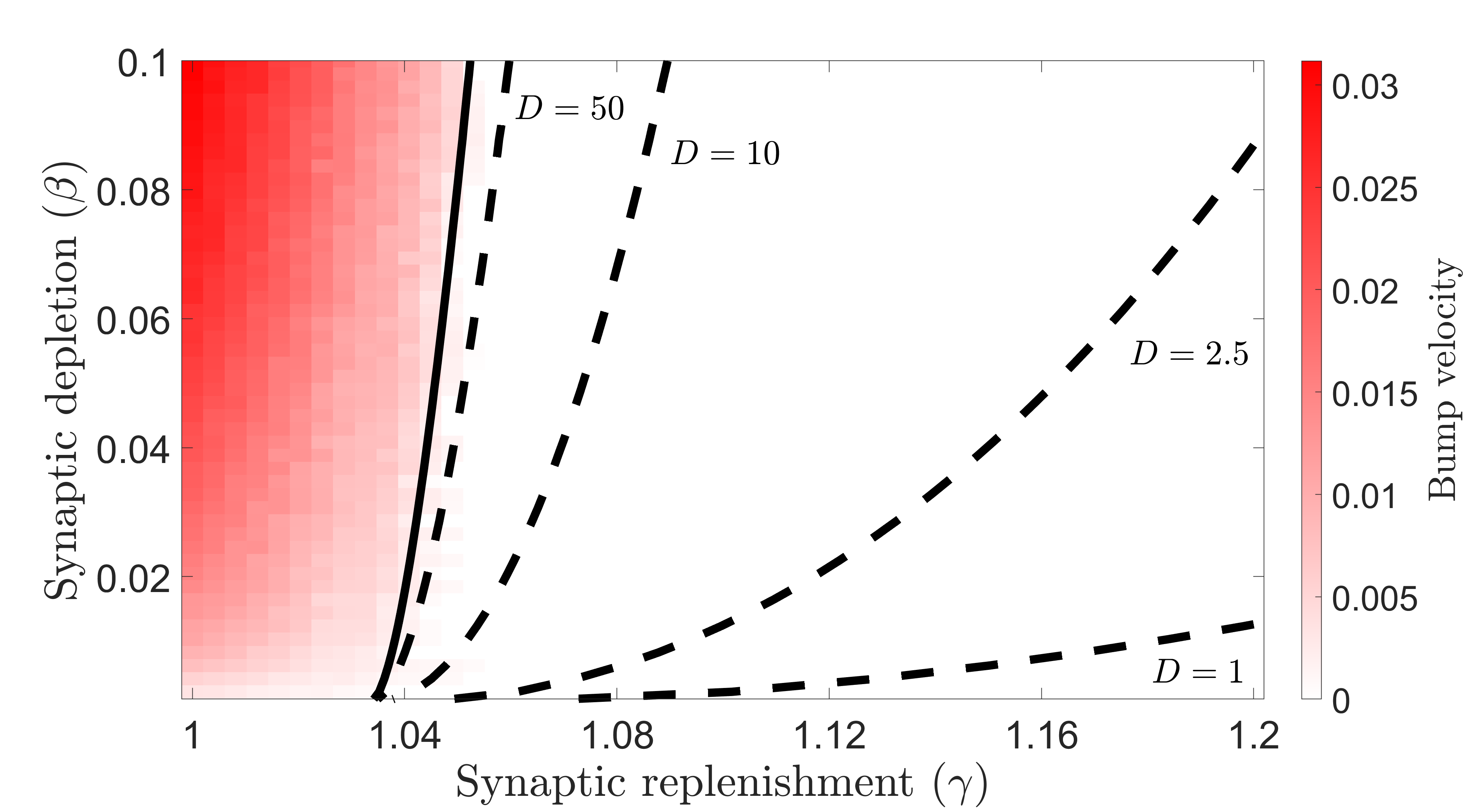}
    \caption{{\bf Stability in the large diffusion limit.} Plots of bump velocity at $T=500$ for simulations of \eqref{eq:anfmodel} with initial conditions specified in \eqref{eq:perturbationIC} for the large diffusion limits $D\to\infty$. Simulations were carried out using $D=1000$ to approximate large diffusion. Dashed curves indicate the boundary where stationary bumps transition from unstable to stable as predicted by section \ref{sec:stability} for $D$ as indicated in the figure. The solid line denotes the boundary predicted by the $D\to\infty$ stability analysis in section \ref{sec:limits}. }
    \label{fig:limits}
\end{figure}

\begin{figure}[t!]
    \centering
\includegraphics[width=\linewidth]{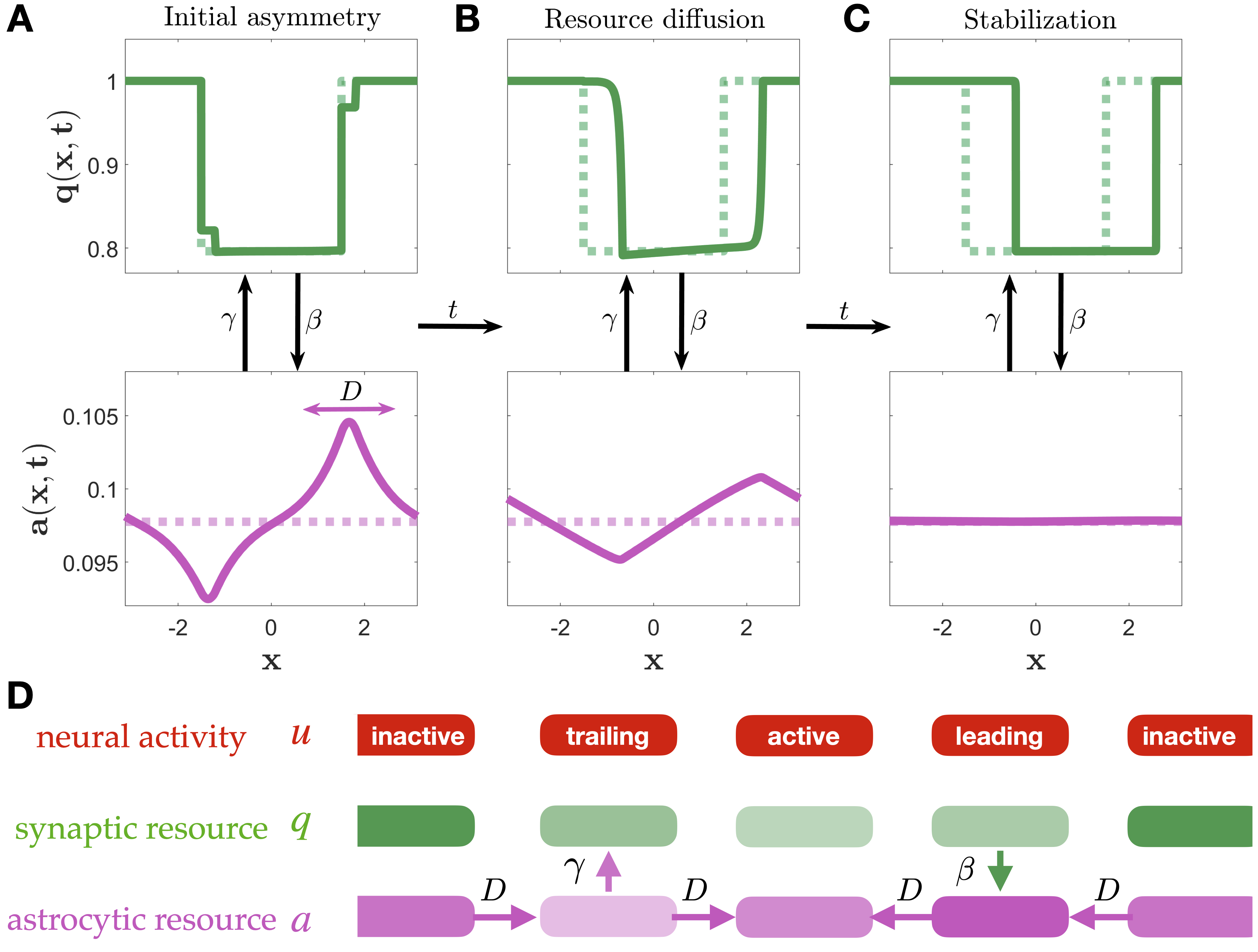}
    \caption{{\bf Mechanism of stabilization.} {\bf A.}~A rightward bump displacement depletes synaptic resources $q$ at the leading edge while astrocytic resources $a$ accumulate there from the newly inactive trailing region, creating a spatial asymmetry that drives further drift. {\bf B.}~Astrocytic diffusion $D$ smooths this asymmetry, redistributing $a$ from regions of high to low concentration and creating a countergradient that opposes the synaptic asymmetry. {\bf C.}~With sufficiently strong $D$ and replenishment rate $\gamma$, the two-step mechanism restores spatial symmetry in both resource pools and a new stationary bump is established. {\bf D.}~Schematic of the stabilization mechanism. Synaptic resource $q$ is depleted at the leading edge ($\beta$) and replenished at the trailing edge ($\gamma$) via astrocytic resource $a$, which diffuses ($D$) inward from the inactive flanks toward the active region.}
    \label{fig:mechanism}
\end{figure}

The Fourier truncation yields a good approximation to the true stability region, though it conservatively predicts a larger region of instability than the exact Evans function analysis (Figure~\ref{fig:phasediagrams}). Consistent with previous work~\cite{Kilpatrick11}, increasing synaptic depletion $\beta$ destabilizes stationary bumps. In contrast to models with purely local resource recovery, this destabilizing effect is partially offset by astrocytic diffusion $D$, which expands the stable region of parameter space and suppresses drift distance monotonically.

\subsection{Mechanistic interpretation and large diffusion limit}
The stabilizing effect of astrocytic diffusion has a natural mechanistic interpretation. As shown in Figure~\ref{fig:limits}, when $\gamma$ is too small the diffusive smoothing cannot be transferred to the synaptic pool rapidly enough to suppress drift, so a threshold replenishment rate is required for diffusion to meaningfully stabilize bumps. Together, sufficiently large diffusion $D$ and replenishment rate $\gamma$ are both necessary for robust stabilization, consistent with the analytic stability conditions derived in Section~\ref{sec:limits}. A rightward shift perturbation creates an asymmetry in the resource pool, with depletion inside the bump and surplus outside, which reinforces further drift in the absence of compensating transport. Astrocytic diffusion counteracts this by transporting resources toward regions of lower concentration, smoothing the asymmetry before it can amplify. Synaptic replenishment then transfers this smoothing back to the synaptic pool, restoring spatial symmetry and stabilizing the bump (Figure~\ref{fig:mechanism}).

\section{Discussion}
\label{sec:discussion}
In this work we introduced an astrocyte-neural field model and established the existence of stationary bump solutions, characterizing their stability by constructing an Evans function whose zeros determine the spectrum of the piecewise-smooth linear stability problem. We have shown that increasing the synaptic depletion rate $\beta$ destabilizes stationary bumps to shift perturbations, causing bumps to drift. In contrast, increasing either astrocytic diffusion $D$ or synaptic replenishment $\gamma$ expands the region of $\beta$ values for which stationary bumps are stable. Notably, the $D=0$ analysis shows that in the absence of spatial resource transport, stability reduces to the simple condition $\gamma>1/2$ independently of $\beta$, confirming that the $\beta$-dependent instability is a consequence of astrocytic diffusion introducing spatial resource gradients, an essential feature of realistic resource trafficking dynamics. The large diffusion limit shows that stabilization of bumps through diffusion is limited by the rate of synaptic replenishment $\gamma$ from the astrocyte resource pool. A two-step stabilization mechanism maintains stable persistent activity: asymmetries in the synaptic resource pool introduced by the shift perturbation are diffusively smoothed in the astrocytic resource pool and then, through synaptic replenishment, symmetry is reestablished in the synaptic resource pool. Without sufficiently strong replenishment, resource asymmetries are created faster than this diffusion and replenishment mechanism can restore spatial symmetry. Thus, strong replenishment and diffusion are both needed for the stabilization of stationary bumps.

Beyond its relevance to biological models of learning and memory, our model belongs to a broader class of coupled nonlocal--local systems in which a spatially nonlocal equation governing a fast variable is coupled to one or more local equations governing slower auxiliary fields. The stability analysis of coherent structures such as bumps, fronts, and pulses in such systems is mathematically nontrivial precisely because the piecewise-smooth linearization couples boundary contributions from the nonlocal term to the local dynamics of the auxiliary fields, as illustrated throughout this paper. Analogous nonlocal--local coupling structures arise in population dynamics, where nonlocal intraspecific competition is coupled to local resource dynamics~\cite{Britton1990}, and in epidemiological models where nonlocal dispersal of infectives is coupled to local susceptible--recovered dynamics~\cite{Medlock2003,Ruan2007}. The techniques developed here for analyzing coherent structure stability in the presence of conserved auxiliary fields may therefore find application beyond neural field theory, particularly in settings where a diffusively coupled reservoir interacts with a nonlocally driven activator.

Several natural extensions remain within the neural field setting. Beyond stationary bumps, neural field models support traveling fronts, pulses, and oscillatory breathers~\cite{Pinto,Folias04,coombes2005bumps,Coombes2003}. 
Synaptic depression alone is known to generate traveling pulses through destabilization of stationary bumps~\cite{Kilpatrick10,bressloff2011two}, and astrocyte-mediated resource dynamics may further enrich this landscape. 
In particular, the interplay between diffusive resource redistribution and local depletion could support novel wave speeds or destabilization patterns not present in purely local models, and understanding how astrocytic diffusion modifies the existence and stability of traveling solutions is a natural next step.

Our analysis assumes a spatially uniform diffusion coefficient $D$. In biological tissue, gap junction coupling between astrocytes is spatially heterogeneous, with regional variation in coupling strength~\cite{Giaume2010,Pannasch2013}. Considering spatially dependent diffusion $D(x)$ would break translation symmetry and create preferential locations for bumps, qualitatively changing the stability landscape. This is directly relevant to the encoding of spatial information in working memory, where bump position encodes a remembered stimulus feature~\cite{Wang}. More broadly, we could develop a model in which the astrocytic network selectively modifies synaptic efficacy as a function of both presynaptic and postsynaptic neuron locations, more completely modifying the effective connectivity kernel $w(x,y)$ in ways that depend on the geometry of glial coupling alongside the synaptic network.

Stationary bumps in neural field models are known to undergo diffusive drift under stochastic perturbations, with drift linked to memory errors in behavioral tasks~\cite{Compte,Wimmer}. A stochastic extension of our  model, in which resource fluctuations drive bump motion, could quantify how astrocytic diffusion reduces the variance of bump motion and improves memory fidelity. Our linear stability analysis of the deterministic system provides the backbone needed for such a systematic weak-noise expansion.

Short-term synaptic facilitation, in which presynaptic activity transiently enhances rather than depletes synaptic efficacy~\cite{tsodyks1998neural}, could directly stabilize bumps by strengthening synaptic feedback within the active region and counteracting the depletion-driven asymmetries that promote drift. Astrocyte-mediated gliotransmission may further modulate this balance, shifting synapses between depressing and facilitating regimes depending on the local glial state~\cite{depitta2016astrocytes}. Incorporating facilitation within the astrocyte-neural field framework could thus reveal how these complementary effects interact with diffusive resource redistribution to shape the emergence and stability of coherent activity patterns.

Finally, the model we have put forth is postulated rather than derived from biophysical principles. An interesting future direction would be to ground it in a stochastic model of activity-dependent glutamate expenditure into the synaptic cleft, diffusion within the extracellular space, and astrocyte uptake and redistribution through the gap junction  network~\cite{lawley2016neurotransmitter,handy2021berg,depitta2016astrocytes}. In the continuum limit, such a framework should recover an astrocyte-neural field model of the form studied here, providing a rigorous link between molecular timescales and the effective rate parameters and diffusion~\cite{Schousboe13,tsodyks1997neural,
abbott1997synaptic}.

Our conserved resource structure connects the model to a broader perspective on memory, in which the finite supply of synaptic resources and their spatial sharing shapes both the capacity and flexibility of memory representations~\cite{BennaFusi2016}. From this perspective, the astrocytic network is an active participant in determining which activity patterns persist and which are displaced, supporting a dynamic competition underlying the brain's ability to continuously and adaptively update its working memory stores.


\section*{Author Contributions}
All authors have made substantial intellectual contributions to the study conception, 
execution, and design of the work. All authors have read and approved the final manuscript.  
In addition, the following contributions occurred:  Conceptualization: Noah Palmer, Daniele Avitabile, Zachary P. Kilpatrick; 
Methodology: Noah Palmer, Heather L. Cihak, Daniele Avitabile, Zachary P. Kilpatrick; Formal analysis and investigation: Noah Palmer; 
Writing - original draft preparation: Noah Palmer; Writing - review and editing: Heather L. Cihak, Daniele Avitabile, Zachary P. Kilpatrick; 
Funding acquisition: Zachary P. Kilpatrick; Supervision: Heather L. Cihak, Daniele Avitabile, Zachary P. Kilpatrick

\section*{Data \& Code Availability} No data was analyzed for this manuscript. Code to reproduce the simulations presented in this manuscript are available from GitHub (\url{https://github.com/noahpalmer/Astrocyte-Neural-Field-Bumps}).

\appendix
\section{Mathematical and biological background}
\label{app:bg}

\subsection*{Bump attractors}
Bump attractors in neural field models are stationary, spatially localized regions of elevated population activity sustained by recurrent synaptic excitation balanced against lateral inhibition~\cite{Amari,Bressloff2012,Pinto,Coombes2003}. Their position on the neural ring encodes a remembered stimulus feature, while models envision drift or diffusion of bump position as generating working memory errors~\cite{Compte,Wimmer}.

\subsection*{Astrocytic resource recycling}
Standard neural field models treat synaptic resource recovery as local and cell-autonomous: each synapse refills its own vesicle pool independently~\cite{tsodyks1997neural,Kilpatrick10}. In biological cortex, however, released glutamate is taken up by surrounding astrocytes and returned to neurons via the glutamate--glutamine cycle~\cite{Schousboe13}. Astrocytes are coupled through gap junctions into a syncytium that transports metabolic substrates laterally across millimeters of tissue~\cite{Giaume2010,Pannasch2013}, meaning a region of sustained firing draws on a spatially extended shared resource pool rather than a purely local one. Our model captures this with conserved total resource split between presynaptic ($q$) and astrocytic ($a$) compartments, with $a$ diffusing spatially to redistribute resources across the tissue.

\subsection*{Linearization and the Evans function}
To assess bump stability, we linearize the system about the stationary solution to derive equations governing small perturbations $\epsilon(\psi, \varphi, \eta)$ and 
seek solutions growing like $e^{\lambda t}$. These equations are valid when $\epsilon \ll 1$. The bump is stable if all eigenvalues $\lambda$ have negative real part. A complication arises because the Heaviside firing rate creates a sharp boundary between active and inactive regions: whether a perturbation expands or contracts this boundary changes the form of the linearized equations, yielding four distinct cases depending on the sign of the perturbation at each bump edge~\cite{coombes2005bumps,bressloff2011two,Cihak2024}.

The Evans function $\mathcal{E}(\lambda)$ is a scalar analytic function whose zeros are exactly the eigenvalues of the linearized operator~\cite{coombes2004evans}, reducing stability analysis to a numerical root-finding problem. In our setting the coupling between neural activity and astrocytic diffusion makes $\mathcal{E}(\lambda)$ transcendental in $\lambda$, so its zeros are located numerically.

\section{Further reading}
\label{app:read}
\begin{itemize}
\item \textit{Neural field theory and bump attractors}: foundational analysis in~\cite{Amari}; modern review in~\cite{Bressloff2012}; broad coverage including Evans functions in~\cite{coombes2014neural}.

\item \textit{Bump attractors and working memory}: spiking network models linking bump dynamics to persistent activity and memory errors in~\cite{Wang,Compte,Wimmer}.

\item \textit{Synaptic depression and resource models}: phenomenological framework and parameter ranges in~\cite{tsodyks1997neural,abbott1997synaptic,tsodyks1998neural}; bump models with depression in~\cite{Kilpatrick10,Cihak2024}.

\item \textit{Astrocyte biology}: glutamate--glutamine cycle in~\cite{Schousboe13,depitta2016astrocytes}; gap junction networks and lateral transport in~\cite{Giaume2010,Pannasch2013}; astrocytes in plasticity and memory in~\cite{Ota2013,Kozachkov2025}.

\item \textit{Evans functions and piecewise-smooth stability}: neural field Evans functions in~\cite{coombes2004evans}; piecewise-smooth stability problems in~\cite{bressloff2011two,bernardo2008piecewise}.
\end{itemize}

\end{document}